\documentclass[journal, twocolumn]{IEEEtran}\IEEEoverridecommandlockouts
\usepackage{amssymb}
\usepackage{amsmath}
\usepackage{amsfonts}
\usepackage{graphicx}
\usepackage{algorithm}
\usepackage{algorithmic}
\usepackage{epstopdf}
\usepackage{cite}
\usepackage{stfloats}
\usepackage{color}
\usepackage{subfigure}
\newtheorem{theorem}{\text{Theorem}}

\newtheorem{lemma}{\text{Lemma}}

\makeatletter
\renewcommand\normalsize{%
   \@setfontsize\normalsize\@xpt\@xiipt
   \abovedisplayskip 0.025\p@ \@plus0.05\p@ \@minus0.125\p@
   \abovedisplayshortskip \z@ \@plus0.075\p@
   \belowdisplayshortskip 0.15\p@ \@plus0.075\p@ \@minus0.025\p@
   \belowdisplayskip \abovedisplayskip
   \let\@listi\@listI}
\makeatother

\begin{document}
\title{\huge Optimal Hybrid Beamforming for Multiuser Massive MIMO Systems With Individual SINR Constraints}
\author{Guangda~Zang,\thanks{
Manuscript received September 1, 2018; revised October 18, 2018; accepted October 20, 2018. This work was supported in part by the National Natural Science Foundation of China under Grant 61401272, Grant 61771309, Grant 61671301, Grant 61420106008, and Grant 61521062, in part by the Shanghai Key Laboratory Funding under Grant STCSM15DZ2270400, in part by the CETC Key Laboratory of Data Link Technology Foundation under Grant CLDL-20162306, and in part by the Medical Engineering Cross Research Foundation of Shanghai Jiao Tong University under Grant YG2017QN47.
The associate editor coordinating the review of this paper and approving it for publication was P. D. Diamantoulakis. \emph{(Corresponding author: Feng Yang.)}

G. Zang is with the Department of Electronic Engineering, Shanghai Jiao Tong University, Shanghai 200240, China, also with
the Department of New Technology of Communication, Shanghai Microwave Research Institute, Shanghai 200331, China, and also with the CETC Key Laboratory of Data Link Technology, China Electronics Technology Group Corporation, Xi’an 710068, China (e-mail: zangguangda@sjtu.edu.cn).

Y. Cui, F. Yang, L. Ding, and H. Liu are with the Department of Electronic Engineering, Shanghai Jiao Tong University, Shanghai 200240, China (e-mail: cuiying@sjtu.edu.cn; yangfeng@sjtu.edu.cn; lhding@sjtu.edu.cn; huiliu@sjtu.edu.cn).

H. V. Cheng is with the Department of Electrical Engineering, Linköping University, 581 83 Linköping, Sweden (e-mail: hei.cheng@liu.se).


}~Ying~Cui,~Hei~Victor~Cheng,~Feng~Yang,~Lianghui~Ding,~and~Hui~Liu
}
\maketitle

\begin{abstract}
In this letter, we consider optimal hybrid beamforming design to minimize the transmission power under individual signal-to-interference-plus-noise ratio~(SINR) constraints in a multiuser massive multiple-input-multiple-output~(MIMO) system.
This results in a challenging non-convex optimization problem.
We consider two cases.
In the case where the number of users is smaller than or equal to that of radio frequency~(RF) chains, we propose a low-complexity method to obtain a globally optimal solution and show that it achieves the same transmission power as an optimal fully-digital beamformer.
In the case where the number of users is larger than that of RF chains,
we propose a low-complexity globally convergent alternating algorithm to obtain a stationary point.
\end{abstract}


\section{Introduction}
\IEEEPARstart{W}{ith}
a large number of antennas deployed in massive multiple-input-multiple-output (MIMO) systems, power consumption and cost of devices increase significantly and may not be affordable for practical implementation.
To address these issues, hybrid analog/digital structure with a reduced number of radio frequency~(RF) chains has been regarded as a promising solution. Analog beamforming refers to the analog operations applied to a signal before being transmitted through antennas, and digital beamforming refers to the baseband signal processing applied to a signal before being sent to RF chains.

Hybrid beamforming technologies have been widely studied in both point-to-point and multiuser massive MIMO systems~\cite{7389996,7387790,8362957}.
It is desirable to consider individual signal-to-interference-plus-noise ratio~(SINR) constraints to guarantee quality of service~(QoS) requirements for different users in multiuser massive MIMO systems.
However, most previous works on multiuser hybrid beamforming design fail to consider individual SINR constraints.
In~\cite{6962946}, the authors consider a non-convex multiuser hybrid beamforming design problem with individual SINR constraints and propose a semidefinite relaxation-based alternating~(SDR-Alt) algorithm to obtain a feasible solution.
In particular, in each iteration, a digital beamforming design problem is solved by computing a semi-closed form solution, and an analog beamforming design problem is solved with complexity $\mathcal{O}(M^{4.5}N^{4.5})$ using standard techniques for semidefinite programming (SDP), where $M$ denotes the number of antennas and $N$ denotes the number of RF chains.
Moreover, most of previous works (e.g.,~\cite{6962946}) focus on the case where the number of users is no greater than that of RF chains, and hence cannot provide meaningful solutions for the emerging massive connectivity applications. To our knowledge, hybrid beamformer optimizations with individual SINR constraints in multiuser massive MIMO systems have not been successfully solved.

In this letter, we consider a multiuser massive MIMO system with $K$ users and $N$ RF chains and assume perfect channel state information (CSI). We study optimal hybrid beamforming design to minimize the transmission power subject to individual SINR constraints.
The resulting challenging non-convex problem is solved in two cases.
In the case of $K\leq N$, we propose a low-complexity method to obtain a globally optimal solution and show that it achieves the same transmission power as an optimal fully-digital beamformer with a reduced number of RF chains, by connecting the original optimization problem to a fully-digital beamforming design problem.
In the case of $K>N$, we propose a low-complexity globally convergent alternating algorithm to obtain a stationary point, based on problem transformation and a penalty method. 
To the best of our knowledge, the proposed solutions are so far the most promising ones in the two cases in terms of computational complexity and theoretical guarantee.
Finally, numerical results show that the proposed solutions have much lower computational complexity than the SDR-Alt algorithm.

\section{System Model and Problem Formulation}
\begin{figure}
\setlength{\abovecaptionskip}{-0.0cm}
\setlength{\belowcaptionskip}{-0.1cm}
\centering
\includegraphics[width=2.2 in]{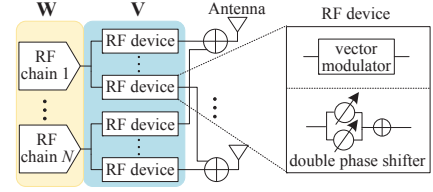}
\caption{Hybrid beamformer structure.}\label{fig_systemmodel}
\end{figure}

Consider a downlink multiuser massive MIMO system with one multi-antenna base station (BS) and $K$ single-antenna users, denoted by $\mathcal{K}\triangleq\{1,\cdots,K\}$.
The BS has $M~(\geq K)$ antennas and $N$ RF chains.
To reduce hardware cost and power consumption, we consider hybrid beamforming with a reduced number of RF chains (i.e., $N<M$).
As illustrated in Fig.~\ref{fig_systemmodel}, we adopt the widely used fully-connected structure, where each RF chain is connected to all $M$ antennas. Thus, the output signal of each antenna can be seen as a linear combination of all RF signals. Let $\mathbf{W}\triangleq[\mathbf{w}_{1},\cdots,\mathbf{w}_{K}]\in{\mathbb{C}}^{N\times K}$ denote the digital beamformer, where $\mathbf{w}_{k}\in{\mathbb{C}}^{N\times 1}$ denotes the digital beamforming vector for user~$k$.
Let $\mathbf{V}\in{\mathbb{C}}^{M\times N}$ denote the analog beamformer.
As in~\cite{VectorModulator,7869042}, we do not impose modulus constraints on the analog beamformer. Note that an analog beamformer without modulus constraints can be implemented using vector modulators~\cite{VectorModulator} or double phase shifter structure~\cite{7869042}.\footnotemark{}\footnotetext[1]{Note that our proposed methods can be extended to the case with modulus constraints by first relaxing the modulus constraints and then projecting the obtained solutions onto the set with modulus constraints.}

We consider a narrowband system and assume a block fading channel model.
Let $\mathbf{g}_k^H\in{\mathbb{C}}^{1\times M}$ denote the channel of user~$k\in\mathcal{K}$, and let $\mathbf{G}\triangleq[\mathbf{g}_1,\cdots,\mathbf{g}_K]^{{H}}\in{\mathbb{C}}^{K\times M}$ denote the channels of the $K$ users, where the superscript $H$ denotes the Hermitian transpose of a matrix. In this letter, we assume perfect CSI at the BS.
The received signal of user~$k$ is given by ${y}_{k}=\mathbf{g}^{H}_{k}\mathbf{V}\mathbf{w}_{k}{s}_{k}+\sum\limits_{i\in\mathcal{K},i\neq k}\mathbf{g}^{H}_{k}\mathbf{V}\mathbf{w}_{i}{s}_{i}+{n}_{k}$,
where ${s}_{k}$ and ${n}_{k}\sim \mathcal{CN}(0, \sigma _{k}^{2})$ denote the transmitted information symbol and the additive Gaussian noise of user $k$, respectively.
We assume that ${s}_{k},k\in\mathcal{K}$ are independent and with zero mean and unit variance. Thus, the transmission power is given by $\Vert \mathbf{V} \mathbf{W}\Vert ^{2}_F$, where $\Vert\cdot\Vert_F$ denotes the Frobenius norm.
To capture the QoS requirement for user~$k\in\mathcal{K}$, we require that the instantaneous SINR of user~$k$ is above a threshold $\eta_k$,
i.e.,
\begin{align}
\frac{\left|\mathbf{g}_k^{H}\mathbf{V}\mathbf{w}_{k}\right|^2}{\sum\limits_{i\in\mathcal{K},i\neq k}{\left|\mathbf{g}_k^{H}\mathbf{V}\mathbf{w}_{i}\right|}^{2}+\sigma_{k}^{2}} \ge \eta _{k},~ k\in\mathcal{K}.\label{eqn:QoS}
\end{align}

Our goal is to optimize the digital beamformer $\mathbf{W}$ and the analog beamformer $\mathbf{V}$ to minimize the transmission power $\Vert \mathbf{V} \mathbf{W}\Vert ^{2}_F$ under the individual SINR constraints in \eqref{eqn:QoS}. Thus, we have the following hybrid beamforming design problem
\begin{equation}\nonumber
\begin{aligned}
\mathcal {P}_{\text{Ori}}: \min _{\mathbf{V},\mathbf{W}}~& \Vert \mathbf{V} \mathbf{W}\Vert ^{2}_F \quad
{\mathrm{s.t.}} ~\eqref{eqn:QoS}.
\end{aligned}
\end{equation}
Problem~$\mathcal {P}_{\text{Ori}}$ is a  challenging non-convex problem.
In Section~\ref{Sec_KleqN} and Section~\ref{sec_K>N}, we shall solve Problem~$\mathcal {P}_{\text{Ori}}$ for two cases, i.e., $K\leq N$ and $K>N$, respectively.

\section{Solution for the Case of $K\leq N$}\label{Sec_KleqN}
In this section, we study the case of $K\leq N$, and obtain a globally optimal solution of the non-convex Problem~$\mathcal {P}_{\text{Ori}}$, by connecting it to a fully-digital beamforming design problem.

First, letting $\mathbf{W}_{D}=\mathbf{V}\mathbf{W}\in\mathbb{C}^{M\times K}$, Problem~$\mathcal {P}_{\text{Ori}}$ can be transformed to the following fully-digital beamforming (with $M$ RF chains) design problem
\begin{align}
\mathcal {P}_{\text{FD}}: \min _{\mathbf{W}_{D}}~& \Vert \mathbf{W}_{D}\Vert ^{2}_F\nonumber \\
{\mathrm{s.t.}} ~&\frac{\left|[\mathbf{g}_k^{H}\mathbf{W}_{D}]_{k}\right|^2}{\sum\limits_{i\in\mathcal{K},i\neq k}{\left|[\mathbf{g}_k^{H}\mathbf{W}_{D}]_{i}\right|}^{2}+\sigma_{k}^{2}}\ge \eta _{k},~  k\in\mathcal{K},\label{eq_FD_constraint}
\end{align}
where $\mathbf{W}_{D}$ can be viewed as the digital beamformer and
$[\ \cdot\ ]_{i}$ denotes the $i$-th element of the argument.
Although Problem~$\mathcal {P}_{\text{FD}}$ is non-convex, it can be solved optimally using several methods, such as the method proposed in~\cite{Wiesel2006} which is based on a semi-closed form solution obtained from KKT conditions and has low computational complexity compared to other methods. Let $\mathbf{W}_{D}^\star$
denote a globally optimal solution of Problem~$\mathcal {P}_{\text{FD}}$.

Next, we construct a globally optimal solution of Problem~$\mathcal {P}_{\text{Ori}}$ based on $\mathbf{W}_{D}^\star$.
Specifically, we randomly generate an $N\times K$ matrix with linearly independent columns, denoted by $\mathbf{W}^\star\in\mathbb{C}^{N\times K}$, and calculate $\mathbf{V}^\star=\mathbf{W}^\star_{D}\left((\mathbf{W}^\star)^H\mathbf{W}^\star\right)^{-1}(\mathbf{W}^\star)^H\in\mathbb{C}^{M\times N}$.
\begin{lemma}\label{lemma_KleqN}
When $K\leq N$,
$(\mathbf{V}^\star, \mathbf{W}^\star)$ is a globally optimal solution of Problem~$\mathcal {P}_{\text{Ori}}$, and $\Vert \mathbf{V}^\star \mathbf{W}^\star\Vert ^{2}_F = \Vert \mathbf{W}^\star_{D}\Vert ^{2}_F$.
\end{lemma}
\begin{IEEEproof}
It is clear that $\mathbf{V}^\star\mathbf{W}^\star=\mathbf{W}_{D}^\star$ and $(\mathbf{V}^\star,\mathbf{W}^\star)$ satisfies~\eqref{eqn:QoS}. Thus, $(\mathbf{V}^\star,\mathbf{W}^\star)$ is a feasible solution of Problem~$\mathcal {P}_{\text{Ori}}$ and achieves the same transmission power as $\mathbf{W}_{D}^\star$. Note that the optimal value of Problem~$\mathcal {P}_{\text{FD}}$ is no greater than that of Problem~$\mathcal {P}_{\text{Ori}}$.
Thus, $(\mathbf{V}^\star,\mathbf{W}^\star)$ is a globally optimal solution of Problem~$\mathcal {P}_{\text{Ori}}$.\end{IEEEproof}

\setlength{\textfloatsep}{0.1cm}
\begin{algorithm}[t]
\caption{Globally Optimal Design for the Case of $K\leq N$}
\label{Alg_Opt}
\begin{algorithmic}[1]
\small{
\STATE Find $\mathbf{W}_{D}^\star$ by solving Problem~$\mathcal {P}_{\text{FD}}$ using the method in~\cite{Wiesel2006};
\STATE Construct $\mathbf{W}^\star\in\mathbb{C}^{N\times K}$ with linearly independent columns;
\STATE Calculate $\mathbf{V}^\star=\mathbf{W}^\star_{D}\left((\mathbf{W}^\star)^H\mathbf{W}^\star\right)^{-1}(\mathbf{W}^\star)^H\in\mathbb{C}^{M\times N}$.
}\normalsize
\end{algorithmic}
\end{algorithm}
\setlength{\floatsep}{0.1cm}

The key steps are summarized in Algorithm~\ref{Alg_Opt}.
To the best of our knowledge, this is the first work providing a globally optimal solution of Problem~$\mathcal {P}_{\text{Ori}}$ and showing that the optimal hybrid beamformer (with at least $K$ RF chains) can achieve the same transmission power as the optimal fully-digital beamformer (with $M$($>N$) RF chains) in the case of $K\leq N$. As Algorithm~\ref{Alg_Opt} requires only computing a semi-closed form solution and some simple matrix operations, it has much lower computational complexity than the SDR-Alt algorithm~\cite{6962946}.

\section{Solution for the Case of $K>N$}\label{sec_K>N}
In this section, we consider the case of $K>N$ and propose a globally convergent alternating algorithm based on a penalty method to obtain a stationary solution of Problem~$\mathcal {P}_{\text{Ori}}$.
\vspace{-0.4cm}
\subsection{Equivalent Problem}
\vspace{-0.1cm}
First, consider the following problem.
\begin{flalign}
\!\!\!\!\mathcal {P}_{\text{Eq}}: &\min _{\mathbf{X}}~ \Vert \mathbf{S}_v \mathbf{X}\mathbf{S}_w\Vert ^{2}_F  \nonumber\\
{\mathrm{s.t.}} &~
\left \|\!\left[\begin{matrix}
(\mathbf{g}_{k}^H\mathbf{S}_v\mathbf{X}\mathbf{S}_w)^H\\
\sigma_k
\end{matrix}\right]\!\right \|_2
\!\!\leq\!\!
\sqrt{\frac{1\!+\!\eta_k}{\eta_k}
}\mathbf{g}_{k}^{H}\mathbf{S}_v\mathbf{X}\mathbf{d}_k,
~\! k\in\mathcal{K},\label{SOCconstraint3}\\
&~\mathbf{g}_{k}^{H}\mathbf{S}_v\mathbf{X}\mathbf{d}_k\geq0,~ k\in\mathcal{K},\label{eq_geq_X}\\
&~\mathbf{X}\succeq\mathbf{0},\label{eq_psd}\\
&~\operatorname{rank}(\mathbf{X})\leq N,\label{eq_rankN}
\end{flalign}
where $\mathbf{S}_v\triangleq[\mathbf{I}_{M\times M},\mathbf{0}_{M\times K}]\in\mathbb{C}^{M\times (M+K)}$, $\mathbf{S}_w\triangleq[\mathbf{0}_{K\times M},\mathbf{I}_{K\times K}]^T\in\mathbb{C}^{(M+K)\times K}$ and $\mathbf{d}_k\in\mathbb{C}^{(M+K)\times 1}$ denotes the vector with the $(M+k)$-th element being $1$ and the rest being $0$.\footnotemark{}\footnotetext[2]{We denote the identity matrix and zero matrix of appropriate size by $\mathbf{I}$ and $\mathbf{0}$, respectively.}
Any feasible solution $\mathbf{X}$ of Problem~$\mathcal {P}_{\text{Eq}}$ can be decomposed as $\mathbf{X}=\mathbf{U}\mathbf{U}^H$ (as $\mathbf{X}$ satisfies the constraints in~\eqref{eq_psd} and~\eqref{eq_rankN}).
We can rewrite $\mathbf{U}$ as $\mathbf{U}=[\mathbf{V}^{H},\mathbf{W}]^{H}$, where $\mathbf{V}\in\mathbb{C}^{M\times N}$ and $\mathbf{W}\in\mathbb{C}^{N\times K}$, i.e., $\mathbf{V}=\mathbf{S}_v\mathbf{U}$ and $\mathbf{W}=\mathbf{U}^H\mathbf{S}_w$. The following result shows the relationship between Problem~$\mathcal {P}_{\text{Ori}}$ and Problem~$\mathcal {P}_{\text{Eq}}$.
\begin{theorem}\label{theorem_equivalent}
If $\mathbf{X}$ is a globally optimal solution of Problem~$\mathcal {P}_\text{Eq}$, $(\mathbf{V},\mathbf{W})$ is a globally optimal solution of Problem~$\mathcal {P}_\text{Ori}$. Furthermore, if
$\mathbf{X}$ is a stationary point of Problem~$\mathcal {P}_\text{Eq}$, $(\mathbf{V},\mathbf{W})$ is a stationary point of Problem~$\mathcal {P}_\text{Ori}$.
\end{theorem}
\begin{IEEEproof}
See Appendix~A.\end{IEEEproof}
\vspace{-0.4cm}
\subsection{Penalty Method}\label{Sec_penalty}
\vspace{-0.1cm}
Based on Theorem~\ref{theorem_equivalent}, we can solve Problem~$\mathcal {P}_{\text{Eq}}$ instead of Problem~$\mathcal {P}_{\text{Ori}}$.
The rank-$N$ constraint in~\eqref{eq_rankN} is non-convex and non-smooth, and hence is hard to deal with.
To address this challenge, instead of~\eqref{eq_rankN}, we consider the following constraint
\begin{equation}\label{eq_eigenvalue}
\begin{matrix}
\operatorname{trace}(\mathbf{X})-\sum_{i=1}^{N}\lambda_i(\mathbf{X})\leq0,\end{matrix}
\end{equation}
where $\lambda_i(\cdot)$ denotes the $i$-th largest eigenvalue of the argument.
As
$\operatorname{trace}(\mathbf{X})\geq\sum_{i=1}^{N}\lambda_i(\mathbf{X})$
holds for any $\mathbf{X}\succeq\mathbf{0}$,~\eqref{eq_eigenvalue} implies $\operatorname{trace}(\mathbf{X})=\sum_{i=1}^{N}\lambda_i(\mathbf{X})$, which means that $\mathbf{X}$ has at most $N$ nonzero eigenvalues, i.e.,~\eqref{eq_rankN} holds. Then we
incorporate \eqref{eq_eigenvalue} as a penalty for violation
and obtain
\begin{align}
&\begin{matrix}
\mathcal {P}_\text{Pen}: \min _{\mathbf{X}}~ \left(\Vert \mathbf{S}_v \mathbf{X}\mathbf{S}_w\Vert ^{2}_F\!+\!\mu (\operatorname{trace}(\mathbf{X})\!-\!\sum_{i=1}^{N}\lambda_i(\mathbf{X}))\right)\end{matrix}\nonumber\\
&\quad\quad\quad\quad~{\mathrm{s.t.}}~\eqref{SOCconstraint3},\eqref{eq_geq_X},\eqref{eq_psd}.\nonumber
\end{align}
Using similar arguments in~\cite{penalty}, we have the following result.
\begin{theorem}\label{Theorem_equivalent_penalty}
There exists $\mu_0~{\in}~(0,+\infty)$ such that for all $\mu ~{>}~ \mu_0$, $\operatorname{trace}(\mathbf{X})-\sum_{i=1}^{N}\lambda_i(\mathbf{X})=0$ and $(\mathbf{V},\mathbf{W})$ is a stationary point of Problem~$\mathcal {P}_\text{Ori}$, where $\mathbf{X}$ is a stationary point of Problem~$\mathcal {P}_\text{Pen}$.
\end{theorem}

Based on Theorem~\ref{Theorem_equivalent_penalty}, we first solve Problem~$\mathcal {P}_\text{Pen}$ for any given $\mu$. Let $\boldsymbol{\Phi}_{M+K,N}~{\triangleq}~\{\mathbf{P}\in\mathbb{S}^{M+K},~\mathbf{0}{\preceq}\mathbf{P}{\preceq}\mathbf{I},~\operatorname{trace}(\mathbf{P})=M{+}K{-}N\}$ denote the convex hull of the rank-$(M{+}K{-}N)$ projection matrices. As 
\begin{equation}
\operatorname{trace}(\mathbf{X})-\sum_{i=1}^{N}\lambda_i(\mathbf{X})=\min_{\mathbf{P}\in\boldsymbol{\Phi}_{M+K,N}}~\operatorname{trace}(\mathbf{P}^T\mathbf{X})
\end{equation}
holds~\cite{nankN}, Problem $\mathcal {P}_\text{Pen}$ can be rewritten as
\begin{align}
\mathcal {P}_{\text{Alt}}: \min _{\mathbf{X}}\min_{\mathbf{P}\in\boldsymbol{\Phi}_{M+K,N}}&~ \left(\Vert \mathbf{S}_v \mathbf{X}\mathbf{S}_w\Vert ^{2}_F+\mu \operatorname{trace}(\mathbf{P}^T\mathbf{X})\right) \nonumber\\
{\mathrm{s.t.}} &~\eqref{SOCconstraint3},~\eqref{eq_geq_X},~\eqref{eq_psd},\nonumber
\end{align}
which can be solved alternatively.
Specifically, let $\mathbf{X}^{(i)}$ denote the estimate of $\mathbf{X}$ at the $i$-th iteration. Then, the estimates of $\mathbf{P}$ and $\mathbf{X}$ at the $(i+1)$-th iteration are updated as
\begin{flalign}
~~\!&\mathbf{P}^{(i+1)}\!=\! \arg\min_{\mathbf{P}\in\boldsymbol{\Phi}_{M+K,N}}~\operatorname{trace}(\mathbf{P}^T\mathbf{X}^{(i)})&\label{eq_alternatingP}
\end{flalign}
\begin{align}
&\mathbf{X}^{(i+1)}\!=\!
\arg\min_{\mathbf{X}}\left(\!\Vert \mathbf{S}_v \mathbf{X}\mathbf{S}_w\Vert ^{2}_F\!+\!\mu \operatorname{trace}((\mathbf{P}^{(i+1)})^T\mathbf{X})\!\!\right)\label{eq_alternatingX}\\
&~\quad\quad\quad\quad\quad{\mathrm{s.t.}}~\eqref{SOCconstraint3},~\eqref{eq_geq_X},~\eqref{eq_psd}.\nonumber
\end{align}
An optimal solution of the convex problem in~\eqref{eq_alternatingP} is given by $\mathbf{P}^{(i+1)}{=}\mathbf{Q}\mathbf{Q}^H$, where $\mathbf{Q}\in\mathbb{C}^{(M+K)\times (M+K-N)}$ is composed of the $M{+}K{-}N$ eigenvectors corresponding to the smallest $M{+}K{-}N$ eigenvalues of $\mathbf{X}^{(i)}$~\cite{nankN} and can be obtained by standard matrix decomposition methods such as singular value decomposition.
The convex SDP problem in~\eqref{eq_alternatingX}
can be solved with complexity $\mathcal{O}((M{+}K)^{4.5})$ using the standard interior-point toolboxes such as SeDuMi. Thus, it is clear that the iterative alternating procedure for given $\mu$ has much lower computational complexity than the SDR-Alt algorithm~\cite{6962946}.
Since $\Vert \mathbf{S}_v \mathbf{X}^{(i)}\mathbf{S}_w\Vert ^{2}_F+\mu \operatorname{trace}(\mathbf{P}^T\mathbf{X}^{(i)})$ is nonnegative and is monotonically non-increasing with $i$,
the iterative alternating procedure for given $\mu$ converges to a limit point.
As the constraint sets of the two problems are disjoint, the limit point is a stationary point of Problem~$\mathcal {P}_{\text{Alt}}$~\cite{stationary}. 
A sufficiently large $\mu$ $(>\mu_0)$ can be found by increasing $\mu$ until $\operatorname{trace}(\mathbf{X})-\sum_{i=1}^{N}\lambda_i(\mathbf{X}){=}0$.

The details are summarized in Algorithm~\ref{Alg_Penalty}.
By Theorem~\ref{Theorem_equivalent_penalty} and by the equivalence between Problem~$\mathcal {P}_{\text{Alt}}$ and Problem~$\mathcal {P}_\text{Pen}$, we know that a stationary point of Problem~$\mathcal {P}_{\text{Ori}}$ can be obtained by Algorithm~\ref{Alg_Penalty}.
As far as we know, this is the first work providing a convergent stationary point of Problem~$\mathcal {P}_{\text{Ori}}$ in the case of $K>N$.

\setlength{\textfloatsep}{0.1cm}
\begin{algorithm}[t]
\caption{Solution for the Case of $K> N$}
\label{Alg_Penalty}
\begin{algorithmic}[1]
\small{
\WHILE {$\operatorname{trace}(\mathbf{X})>\sum_{i=1}^{N}\lambda_i(\mathbf{X})$}
\STATE {construct $\mathbf{X}^{(0)}$ with random values and set $i:=0$}
\REPEAT
\STATE Obtain $\mathbf{P}^{(i+1)}$ by solving the problem in~\eqref{eq_alternatingP};
\STATE Obtain $\mathbf{X}^{(i+1)}$ by solving the problem in~\eqref{eq_alternatingX};
\STATE $i \gets i+1$;
\UNTIL{convergence criterion is met;}
\STATE $\mu:=2\mu$;
\ENDWHILE
}\normalsize
\end{algorithmic}
\end{algorithm}
\setlength{\floatsep}{0.1cm}

\section{Numerical Results}\label{Sec_simulation}
In this section, we provide numerical results to illustrate the performance of Algorithm~\ref{Alg_Opt} and Algorithm~\ref{Alg_Penalty}.
In the simulations, the one-ring channel model is used by setting the angular spread as~$\Delta =15^{\circ}$ and assuming the azimuth angle of arrival for user~$k$ as $\theta_k =-180^{\circ}+\Delta +(k-1)\frac{360^{\circ}}{K}$. We choose $\eta_k=\sqrt{2}-1$ and $\sigma_{k}^{2}=1$.
We consider four baselines for comparison.
The first baseline is the hybrid beamformer obtained using the SDR-Alt algorithm in~\cite{6962946} for solving Problem~$\mathcal {P}_{\text{Ori}}$.
The other three baselines are three typical fully-digital beamformers ($N=M$), i.e., the optimal solution $\mathbf{W}_{D}^\star$ of Problem~$\mathcal {P}_{\text{FD}}$ (optimal fully-digital beamformer), fully-digital beamformer based on zero-forcing~(ZF) and fully-digital beamformer based on maximum-ratio-transmission~(MRT), which satisfy the SINR constraints in~\eqref{eq_FD_constraint}.
In evaluating the two proposed algorithms and the SDR-Alt algorithm in~\cite{6962946}, we use the same convergence criterion; we generate 30 random channels (same for all schemes), and show the mean and standard deviation (cf. vertical bar at each point) of the performance. We compare the normalized average power consumption which is unit-less.

\begin{figure}
\setlength{\abovecaptionskip}{-1pt}
\setlength{\belowcaptionskip}{0pt}
\centering
\includegraphics[width=2.3 in]{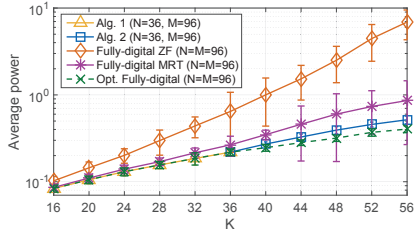}
\caption{Average power versus $K$ at $N=36$ and $M=96$.}\label{fig_versusK}
\end{figure}

\begin{figure}
\setlength{\abovecaptionskip}{-1pt}
\setlength{\belowcaptionskip}{-0pt}
\centering
\subfigure[Average power versus $M$.]{
\label{comparison_power}
\includegraphics[width=1.4 in]{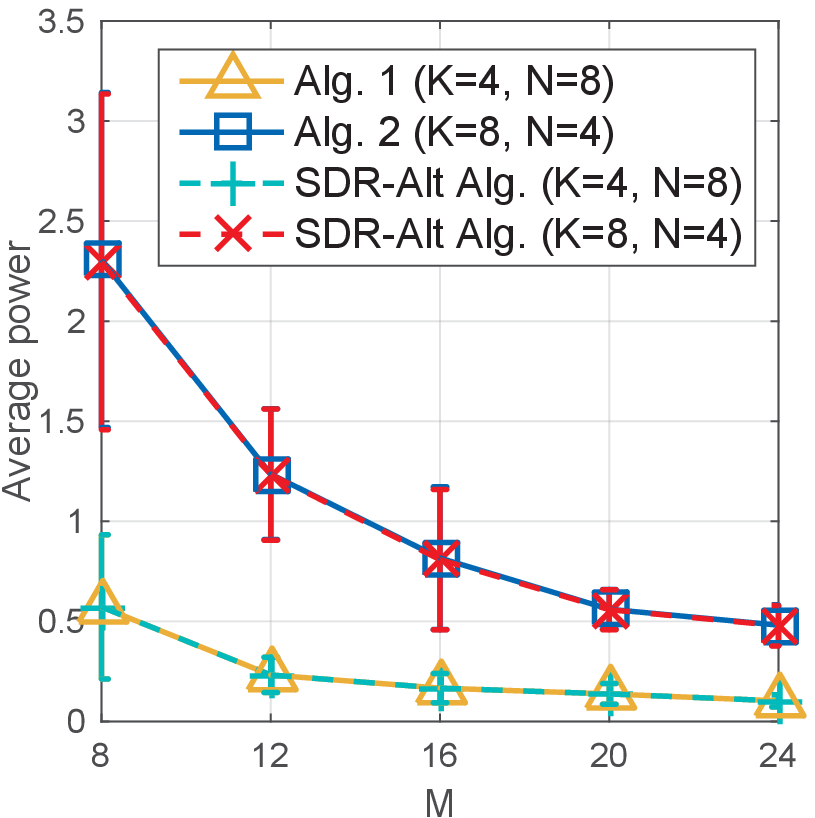}} 
\subfigure[Simulation time versus $M$.]{
\label{comparison_time}
\includegraphics[width=1.4 in]{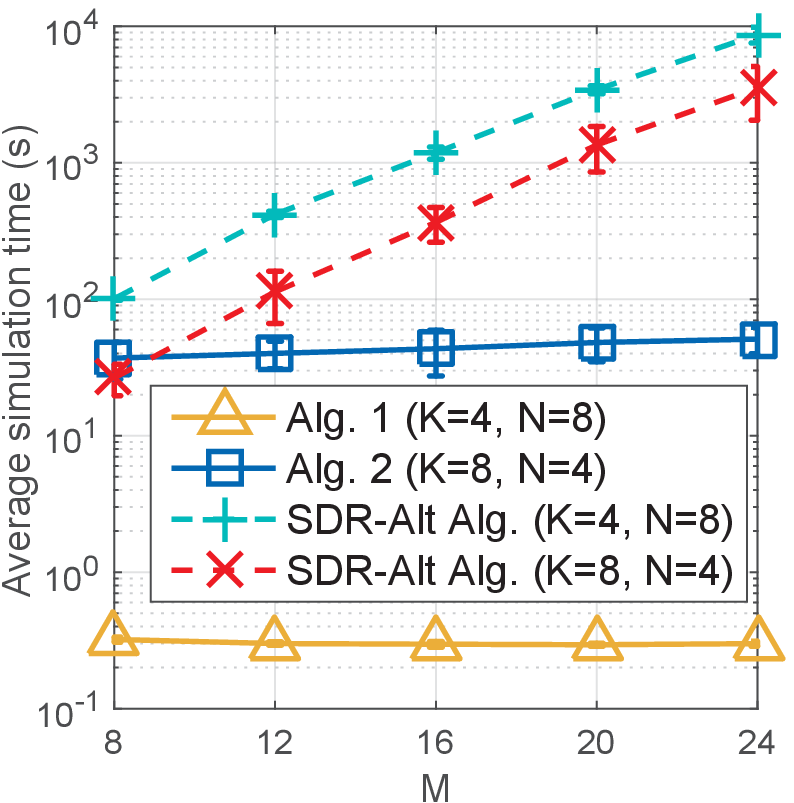}}
\caption{Average power and simulation time.}\label{fig_complexity}
\end{figure}

Fig.~\ref{fig_versusK} illustrates the average power versus the number of users $K$.
We can observe that, in the case of $K{\leq} N$, Algorithm~\ref{Alg_Opt} achieves the same average power as the optimal fully-digital beamformer.
In the case of $K{>}N$, Algorithm~\ref{Alg_Penalty} outperforms the fully-digital beamformers based on ZF and MRT, and achieves similar average power compared to the optimal fully-digital beamformer.
These indicate that hybrid beamforming can achieve most of beamforming performance with reduced hardware cost.
In Fig.~\ref{fig_versusK}, we do not provide results for the SDR-Alt algorithm, as its computational complexity at $N{=}36$ and $M{=}96$ is not acceptable.
In Fig.~\ref{fig_complexity}, we compare the average power and simulation time (reflecting computational complexity) of the proposed algorithms and the SDR-Alt algorithm at small $N$ and $M$.
The proposed algorithms achieve the same average power as the SDR-Alt algorithm with much lower computational complexity. In addition, the computational complexity of the proposed algorithms almost does not change with $M$, while the computational complexity of the SDR-Alt algorithm increases dramatically with $M$. Thus, Fig.~\ref{fig_versusK} and Fig.~\ref{fig_complexity} demonstrate the advantages of the proposed algorithms over the SDR-Alt algorithm.

\section{Conclusion}
In this letter, we considered the optimal hybrid beamforming design in a multiuser massive MIMO system to minimize the total transmission power under individual SINR constraints. By exploring structural properties of the problem, we proposed two low-complexity algorithmic solutions to solve the challenging non-convex problem in two cases depending on the number of users and the number of RF chains.
The computational complexity of the proposed algorithms is dramatically reduced compared to the existing SDR-Alt algorithm.

\appendices
\renewcommand{\thesectiondis}[2]{\Alph{section}. }
\section{Proof of Theorem~1}\label{Sec_App_Proof}

First, it can be verified that multiplying a feasible point $\mathbf{V}\mathbf{W}$ of Problem~$\mathcal {P}_{\text{Ori}}$ on the right by a diagonal phase scaling $\operatorname{diag}(e^{j\phi_i})$, where $\phi_i,i=1, \ldots,K$ are arbitrary phase values and $\operatorname{diag}(x_i)$ denotes a diagonal matrix with $x_i$ being the $i$-th diagonal element, the feasibility and objective value of Problem $\mathcal {P}_{\text{Ori}}$ do not change.
If $\mathbf{V}^\star\mathbf{W}^\star$ is an optimal solution, then $\mathbf{V}^\star\mathbf{W}^\star\operatorname{diag}(e^{j\phi_i})$ is also an optimal solution.
Thus, we can restrict the $k$-th diagonal element of $\mathbf{G}\mathbf{V}\mathbf{W}$, i.e., $\mathbf{g}_{k}^{H}\mathbf{V}\mathbf{w}_{k}$, to the non-negative real domain and impose
\begin{equation}\label{eq_geq_ori}
\mathbf{g}_{k}^{H}\mathbf{V}\mathbf{w}_{k}\geq0,~ k\in\mathcal{K}.
\end{equation}
By~\eqref{eqn:QoS}, we have
\begin{align}
\begin{matrix}
\sum\limits_{i\in\mathcal{K},i\neq k}{\left|\mathbf{g}_k^{H}\mathbf{V}\mathbf{w}_{i}\right|}^{2}{+}\sigma_{k}^{2} \leq \frac{1}{\eta _{k}}\left|\mathbf{g}_k^{H}\mathbf{V}\mathbf{w}_{k}\right|^2,~ k\in\mathcal{K},\nonumber
\end{matrix}
\end{align}
\begin{align}
\begin{matrix}
{\Rightarrow}~\sum\limits_{i\in\mathcal{K}}{\left|\mathbf{g}_k^{H}\mathbf{V}\mathbf{w}_{i}\right|}^{2}{+}\sigma_{k}^{2} \leq (1{+}\frac{1}{\eta _{k}})\left|\mathbf{g}_k^{H}\mathbf{V}\mathbf{w}_{k}\right|^2,~ k\in\mathcal{K},\\
\end{matrix}\nonumber
\end{align}
\begin{align}
\begin{matrix}
{\Rightarrow}\left\| \left[\begin{matrix}
(\mathbf{g}_{k}^{H}\mathbf{V}\mathbf{W})^{H}\\
\sigma_k
\end{matrix}\right]\right \|^2_2
\leq
(1{+}\frac{1}{\eta _{k}})
\left|\mathbf{g}_{k}^{H}\mathbf{V}\mathbf{w}_{k}\right|^2,
~ k\in\mathcal{K}.
\end{matrix}\nonumber
\end{align}
Taking square root on both sides of the above inequality and by~\eqref{eq_geq_ori}, we have
\begin{equation}\label{SOCconstraint}
\left\| \left[\begin{matrix}
(\mathbf{g}_{k}^{H}\mathbf{V}\mathbf{W})^{H}\\
\sigma_k
\end{matrix}\right]\right \|_2
\leq
\sqrt{\frac{1\!+\!\eta_k}{\eta_k}
}\mathbf{g}_{k}^{H}\mathbf{V}\mathbf{w}_{k},
~ k\in\mathcal{K},
\end{equation}
Next, letting $\mathbf{U}\triangleq[\mathbf{V}^{H},\mathbf{W}]^{H}\in{\mathbb{C}}^{(M+K)\times N}$, we have $\mathbf{V}=\mathbf{S}_v\mathbf{U}$, $\mathbf{W}=\mathbf{U}^H\mathbf{S}_w$ and $\mathbf{w}_k=\mathbf{U}^H\mathbf{d}_k$.
Thus,~\eqref{eq_geq_ori} and~\eqref{SOCconstraint} can be rewritten as
\begin{flalign}
&\quad\quad\quad\quad\quad\quad\mathbf{g}_{k}^{H}\mathbf{S}_v\mathbf{U}\mathbf{U}^H\mathbf{d}_k\geq0,~ k\in\mathcal{K}.\label{eq_U_geq}\\
&\!\!\!\!\!\left \|\!\left[\begin{matrix}\!
(\mathbf{g}_{k}^{H}\mathbf{S}_v\!\mathbf{U}\mathbf{U}^H\!\mathbf{S}_w)^{H}\\
\sigma_k
\end{matrix}\right]\!\right \|_2
\!\!\leq\!
\sqrt{\frac{1\!+\!\eta_k}{\eta_k}
}\mathbf{g}_{k}^{H}\mathbf{S}_v\mathbf{U}\mathbf{U}^H\mathbf{d}_k,
 k\!\in\!\mathcal{K}.\label{SOCconstraint2}
\end{flalign}
Then Problem~$\mathcal {P}_{\text{Ori}}$ can be equivalently transformed to
\begin{align}\nonumber
\mathcal {P}_\text{Re}:\min _{\mathbf{U}}&~ \Vert \mathbf{S}_v \mathbf{U}\mathbf{U}^H\mathbf{S}_w\Vert ^{2}_F \\
\mathrm{s.t.}&~\eqref{eq_U_geq},~\eqref{SOCconstraint2}.\nonumber
\end{align}
Note that $\mathbf{X}$ can be rewritten as $\mathbf{X}=\mathbf{U}\mathbf{U}^H\in\mathbb{C}^{(M+K)\times (M+K)}$ for some $\mathbf{U}$ if and only if $\mathbf{X}$ satisfies constraints~\eqref{eq_psd} and~\eqref{eq_rankN}. Thus, if $\mathbf{X}$ is a globally  optimal solution of Problem~$\mathcal {P}_\text{Eq}$, $(\mathbf{V},\mathbf{W})$ is a globally optimal solution of Problem~$\mathcal {P}_\text{Ori}$.
Furthermore, it can be verified that if $\mathbf{X}$ satisfies the KKT system of Problem~$\mathcal {P}_\text{Eq}$, $\mathbf{U}$ also satisfies the KKT system of Problem~$\mathcal {P}_\text{Re}$. Thus, if $\mathbf{X}$ is a stationary point of Problem~$\mathcal {P}_\text{Eq}$, $\mathbf{U}$ is a stationary point of Problem~$\mathcal {P}_\text{Re}$.
Besides, by similar calculations provided in~\cite[Proposition 3]{Wiesel2006}, if $\mathbf{U}$ satisfies the KKT system of Problem~$\mathcal {P}_{\text{Re}}$, $(\mathbf{V},\mathbf{W})$  satisfies the KKT system of Problem $\mathcal {P}_{\text{Ori}}$. Thus, if $\mathbf{U}$ is a stationary point of Problem~$\mathcal {P}_\text{Re}$, $(\mathbf{V},\mathbf{W})$ is a stationary point of Problem~$\mathcal {P}_\text{Ori}$. Therefore, if $\mathbf{X}$ is a stationary point of Problem~$\mathcal {P}_\text{Eq}$, $(\mathbf{V},\mathbf{W})$ is a stationary point of Problem~$\mathcal {P}_\text{Ori}$.

%
%

\ifCLASSOPTIONcaptionsoff
  \newpage
\fi

\bibliographystyle{IEEEtran}
\bibliography{yang_WCL2018-1093_R1_bib}

\end{document}